\documentstyle[aps,preprint,psfig]{revtex}

\newcommand{\PRA}[1]{Phys.\ Rev.\ A {\bf #1}}

\newcommand{\PRE}[1]{Phys.\ Rev.\ E {\bf #1}}
\newcommand{\PRL}[1]{Phys.\ Rev.\ Lett.\ {\bf #1}}

\newcommand{\PLA}[1]{Phys.\ Lett.\ A {\bf #1}}

\newcommand{\PA}[1]{Physica A {\bf #1}}
\newcommand{\JSP}[1]{J.\ Stat.\ Phys. {\bf #1}}
\newcommand{\PS}[1]{Physica.\ Scripta. {\bf #1}}
\newcommand{\EL}[1]{Europhys.\ Lett. {\bf #1}}
\begin{document}


\draft

\title{ Disorder-induced phase transition in a one-dimensional model of 
rice pile}

\author{M. BENGRINE$^{1,2}$, A. BENYOUSSEF$^{2}$, 
F. MHIRECH\cite{em1}$^{1,2}$ and S. D. ZHANG\cite{em2}$^{1,3,4}$}
\address{$^1$ International Centre For Theoretical Physics, Trieste, Italy\\
$^2$  Laboratoire de Magnetisme et de Physique des Hautes Energies 
, Departement de Physique, Faculte des Sciences, Rabat, Morocco\\
$^3$Institute of Low Energy Nuclear Physics,
Beijing Normal University, Beijing 100875, China\\
$^4$ Beijing Radiation Center, Beijing 100088\\
}

\date{\today}

\maketitle

\begin{abstract}
We propose a one-dimensional rice-pile model which connects the 1D BTW 
sandpile model ( Phys. Rev. A 38, 364 (1988) ) and the Oslo rice-pile 
model ( Phys. Rev. lett. 77, 107 (1997) ) in a continuous manner. 
We found that for a sufficiently large system, there is a sharp transition 
between the trivial critical behaviour of the 1D BTW model and the self-organized 
critical (SOC) behaviour. When there is SOC , the model belongs to a known 
universality class with the avalanche exponent $\tau=1.53$.

\end{abstract}

\maketitle


Bak et al proposed "self-organized criticality" (SOC) as a framework to 
understand the dynamics of driven, dissipative systems~\cite{BTW}. 
These systems, via self-organization, reach the steady state which is 
characterized by a power-law distribution of the sizes of avalanches. 
The authors of Ref.~\cite{BTW} originally used a cellular automata, now referred to 
as the BTW sandpile model, to illustrate their ideas. Moreover the SOC behavior was 
also found in some biological~\cite{btw2} and economical~\cite{Ja} systems. 
Different variations of the BTW sandpile were proposed and studied~\cite{Kad,dhar1,Manna1,chau1,priez1}. 
The behaviors of these sandpiles are somewhat different depending on whether the rules of 
evolution are based on the absolute sand heights of the pile~\cite{Manna1,chau1,priez1}, 
the local slopes~\cite{Kad,Manna1}, or the Laplacians of the sand 
height function~\cite{Manna1}. 
Besides the 
numerical simulations many different methods were also used to treat 
the SOC problems. 
Dynamical mean field theory~\cite{ves3} gives a unified description 
of some stochastic SOC systems including the BTW sandpile model and the 
forest fire model~\cite{btw4}. 
Langevin-type approaches~\cite{guil1} have been used on a phenomenological 
basis. Furthermore, a real space renormalization group method~\cite{ves1} 
provided good estimates of the avalanche exponents. 
Finally, it has been shown~\cite{dhar1} that a large class of sandpiles  
were Abelian and this property leads to a particularly simple equiprobable 
partitioning in configuration space that allows to extract some exact results. 
Numerical simulations of high dimensional BTW model\cite{Lubeck} were recently performed to determine the 
upper critical dimension where the avalanche distributions are characterized by the mean-field 
exponents. 
The idea of SOC also stimulated much interest in the 
granular matter and some experiments\cite{Jaeger} were done to investigated whether 
real sandpile display SOC behavior. 
In order to make comparison with theories and models, a group of researchers 
in Oslo did experiments on 
real rice piles~\cite{Frette} and showed that under some conditions a real rice pile 
displays SOC behaviors. 
In fact, for grains with a large aspect ratio the system self-organizes 
into a critical state. They explained this result with the increased friction and 
different packing possibilities. 
By measuring the transit time after the pile has reached the 
stationary critical state, they found that the distribution of the transit 
times follows the form:
\begin {equation}
P(T,L)=L^{-\beta}F(T/L^{\nu}),
\label{P(T,L)}
\end {equation}
where $T$ is the transit time, $L$ is the system size. The scaling function 
$F(x)$ is constant for small $x$ and decays as power law with a slope 
$\alpha=2.4\pm0.2$ for larger $x$. 

To take into account the changes in the local slopes observed in the rice pile 
experiment, 
Christensen {\em et al} proposed a rice-pile model, hereafter called Oslo 
model~\cite{Chris,Amaral,Paczuski}, where the critical slope for 
each site is a  dynamical variable. 
The Oslo model is based on a linear array of cells labelled by $i$, where 
$i=1,2,...,L$, and an integer variable $h(i)$ assigned to each of them, with a 
wall at $i=0$ and an open boundary at $i=L+1$. Here $h(i)$ is called the local height 
of the rice pile at site $i$. The local slope assigned to each site $i$ is defined 
as $z(i)=h(i)-h(i+1)$ for $i=1,2,....,L$.
Initially, the system is empty, i.e., $h(i)=0$ $\forall i$. The dynamics of 
the model consists of  deposition and relaxation, and the relaxation process 
is considered to be fast compared to the deposition time scale. At each time 
step one grain is added to a column $i$ which increase the height of $i$ by 1, 
i.e., $h(i) \rightarrow h(i)+1$ . 
With the dropping of rice grains, a rice pile is built up. 
Whenever there is active column, i.e., $z(i)>z^{c}(i)$, where $z^{c}(i)$ is a 
slope threshold, one grain of rice will be transferred from this column to its 
right neighbor, $h(i) \rightarrow h(i)-1$ and $h(i+1) \rightarrow h(i+1) +1$, 
and all the unstable sites topple in parallel. 
The critical slope $z^c$ of a site remains unchanged if 
the site is 
stable but assumes a new value $1$ or $2$ randomly every time a 
rice grain on this site has toppled. 
The toppling of one or more sites is called an avalanche event, and during 
the avalanche no grains are added to the pile. The avalanche stops when the 
system reaches a stable state with $z(i)\leq z^{c}(i)$ $\forall i$. The 
internal randomness in the critical slopes makes the Oslo model different from 
the 1D BTW model. 
For the Oslo model, with  arbitrary initial conditions, the system reaches a 
stationary state where the avalanche sizes are power-law distributed, while 
for the 1D BTW model the 
size distribution of avalanche is generally not power-law. Since the only difference 
between the 1D BTW model and the Oslo model is the presence of internal disorder in the 
latter, it is natural to consider that the criticality in the Oslo model is 
induced by the disorder ( randomness in the critical slopes). 
One motivation of this letter is to 
investigate the  transition between the 1D BTW model and the Oslo model. 

We modify the Oslo model as follows: When $z(i)\leq z^{c}(i)$ $\forall i$, 
the pile is stable and there isn't diffusion of particles. If, at site $i$, 
$z(i)>z^{c}(i)$, where $z^{c}(i)$ takes randomly a value $1$ or $2$, then 
this site 
topples with a probability which depends on its slope $z(i)$, namely: if 
$1<z(i)\leq2$, the site $i$ topples with a probability $p_{1}$ and if $z(i)>2$, 
it topples with probability $p_{2}$. Notice that if we set $p_{1}=0$ and 
$p_{2}=1$, the model becomes the BTW model with the critical slope $z_{c}=2$, 
and if we set $p_{1}=p_{2}=1$, it is just the Oslo model.     
In a previous paper~\cite{zhang}, one of us generalized the 
Oslo model in a different way, in which the critical local slope can assume $r$ 
different values, from 1 to $r$, where $r$ is an integer. It is possible 
to show that the exponents for the avalanche-size and transit-time 
distributions are insensitive to the level of medium disorder 
(different values of $r>1$) when the grain is dropped at a 
fixed position. So in the present model we will only consider the case 
$r=2$, i.e., the critical slope takes randomly the value 1 or 2. 
In this study, we will restrict ourselves to the case where a grain is added 
to the site i=1. 
When a grain is dropped on 
the left-end site of the pile, it may make the site unstable. The site topples 
and transfers a grain of rice to its right neighbour and so on. And in this way 
avalanches occur. As in the 
literature, we define the size of an avalanche as the number of toppling. 
In this letter, we will take $p_{2}$ equal to $1$ and $p_{1}\leq p_{2}$ because 
higher is the slope, higher is the jump probability.
So by varying $p_1$ from $0$ to $1$, we can change the model from the 1D 
BTW sandpile model to the Oslo model in a continuous manner. 

We have performed extensive numerical simulations and investigated the effect 
of $p_1$ on the behaviors of the model. 

\begin{figure}
\centerline{\psfig{file=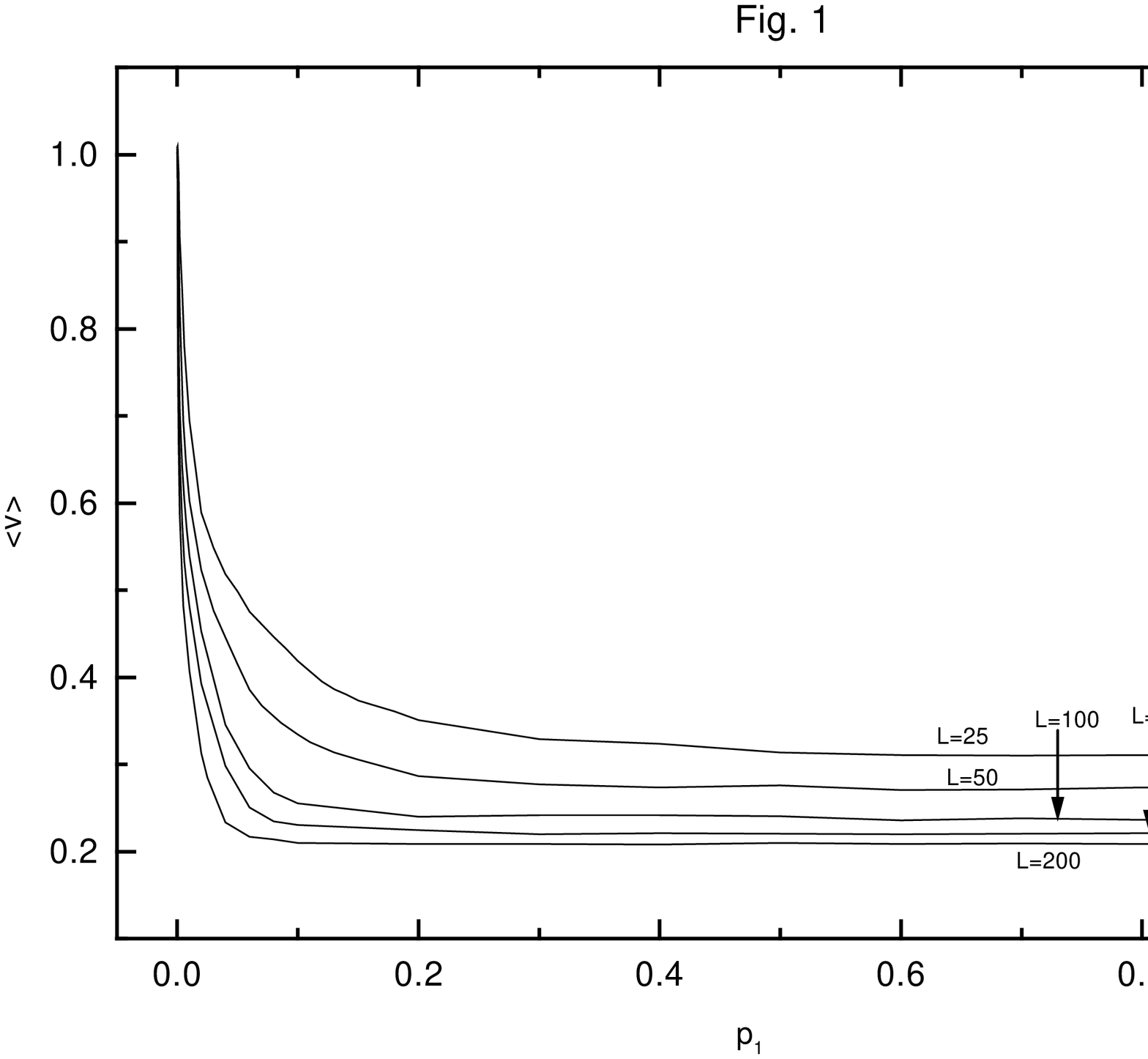,width=3in,angle=0}}
\caption{
The average velocity $<v>$, defined in the text, as a function of the 
probability $p_{1}$ for several values of the system size.
}
\label{fig<v>p1}
\end{figure}

Let us first study the transport properties of the model. 
The transit time of a grain is defined as the time it spent in the pile, and the 
time is measured in the unit of additions of grains.  
When a grain slips out of the pile, we can measure its transit time $T=T_{out}-T_{in}$, 
where $T_{out}$ and $T_{in}$ denote the input and output time of the grain. 
For the case 
$p_1=0$ (the 1D BTW model) when the stationary state is reached every 
newly-added 
grain will slip out of the pile instantly, thus the transit time 
is $T=0$, and the average transportation velocity 
of grains, defined as $<v>=L/<T>$, is infinite for this case. On the other 
hand, for the case $p_1=1$ (the Oslo model) previous studies show that 
the average transportation velocity scales with the system size as 
$<v>\propto L^{-\gamma}$ with $\gamma =0.30 \pm 0.10$, indicating that $<v>\rightarrow 0$ in the limit of 
infinite system size. It would be interesting to see 
how $<v>$ changes from $\infty$ to $0$ when $p_1$ is varied from $0$ to 
$1$. In fact, we found that there is sharp transition at $p_1=0$. 
The general behavior of $<v>$ as a function of $p_1$ is shown in 
Fig.~\ref{fig<v>p1}.  One can see from Fig.~\ref{fig<v>p1} that for a given 
system size $L$, the velocity 
$<v>$ becomes constant when $p_1$ is larger than some value, say $p_1^c$, 
where $p_1^c$ itself is dependent on the system size and becomes closer to $0$ 
when $L$ becomes larger. The numerical results make us to consider that 
$p_1^c\rightarrow 0$ as $L\rightarrow \infty$. 
. When $p_1 \rightarrow 0$, $<v>=1$, independent of the system size. 
So there is a sharp transition from $<v>=\infty$ for $p_1=0$ to $<v>=1$ for 
$p_1=0^+$.  

At first think, the sharp transition may be surprising. But it can be understood 
by the following argument.  It is clear that when $p_1$ is exactly $0$, 
no newly-added grain will stay 
in the pile as long the stationary state is reached. So $T=0$ for every 
grain, and hence $<T>=0$. For $p_1=0^+$ however, the situation is 
different, and there is the possibility that some grains will be buried in the 
surface layer of the pile. These grains will stay 
in the pile for a very long time. Once they slip out of the pile, these 
grains, although very few in number, will  
make a significant contribution to $<T>$ since their transit times are 
extremely large. It is the existence of these grains that makes 
$<T>$ assume finite value for $p_1=0^+$. Thus the sharp transition here is 
induced by the tiny disorder.
Between $p_1=0^+$ and $p_1=p_1^c$, there is crossover behavior of $<v>$, 
which is due to finite size effects. Since we expect $p_1^c \rightarrow 0$ when 
$L\rightarrow \infty$, we can also expect that for infinite system the 
transition takes place at $p_1=0$ form $<v>=\infty$ to $<v>=0$.
 
\begin{figure}
\centerline{\psfig{file=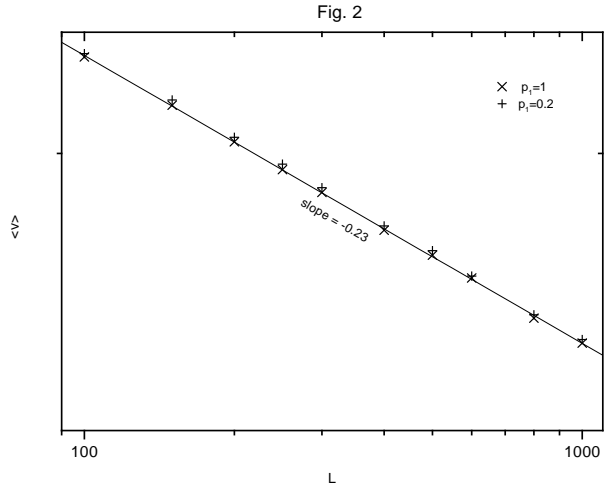,width=3in,angle=0}}
\caption{The average velocity $<v>$ as a function of the system size $L$, for 
two values of the probability $p_{1}$.
}
\label{fig<v>L}
\end{figure}

In what follows we shall show that the critical behaviors of the model 
belong to the same universality class of the Oslo model When $p_1$ is larger than $p_1^c$. 
In Fig.~\ref{fig<v>L}, we plot the average 
velocity as a function of the system size for different values of $p_{1}$ 
greater than $p_1^c$. It is clear that for large enough system 
($L>100$), the average velocity $<v>$ scales as $L^{-\gamma}$, where $\gamma \approx 0.23$.
Therefore the average velocity decreases with the system size, which is 
due to the increase in the active zone depth with system size, 
as explained  by Christensen {\em et al}~\cite{Chris}. 

\begin{figure}
\centerline{\psfig{file=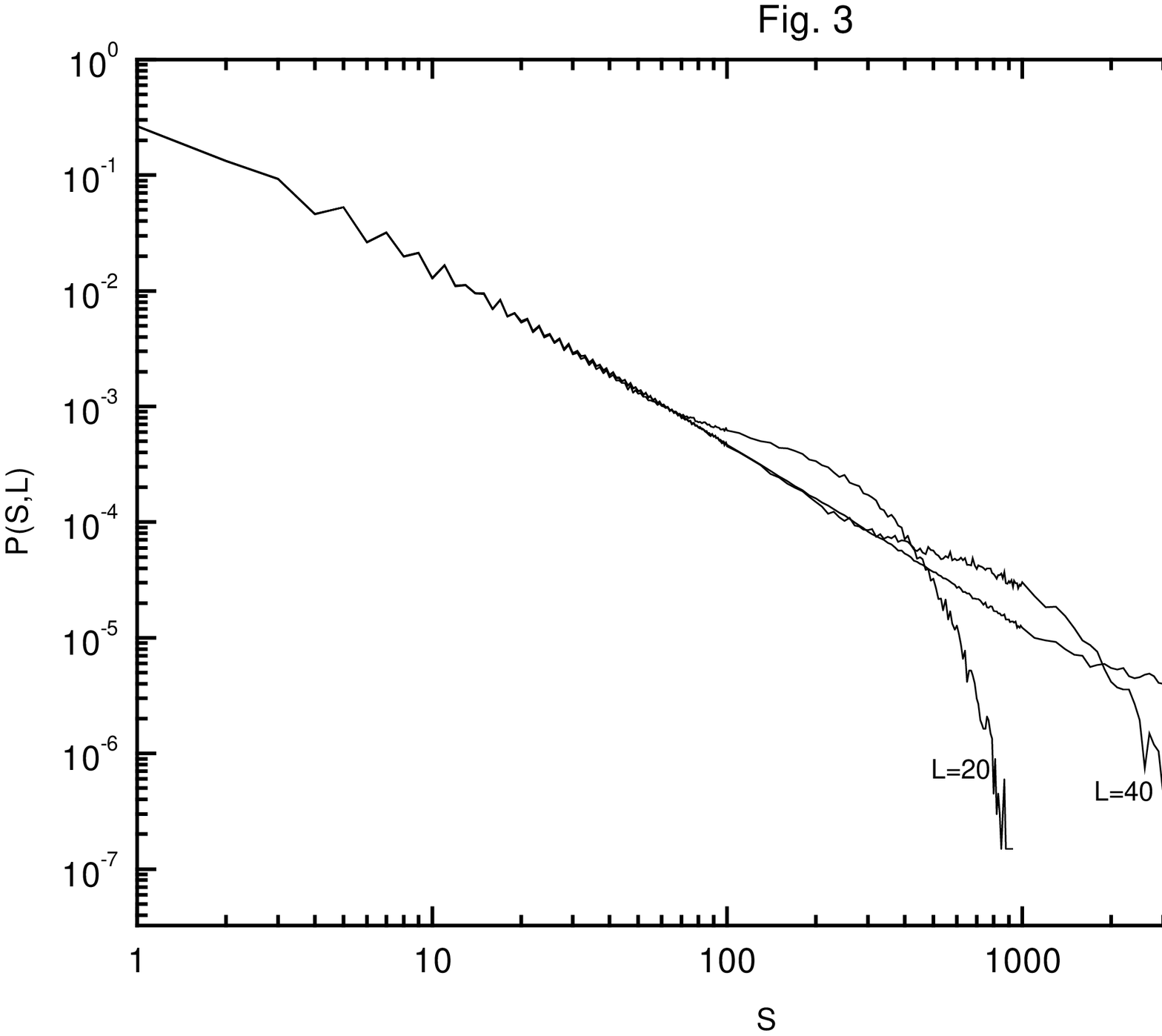,width=3in,angle=0}}
\centerline{\psfig{file=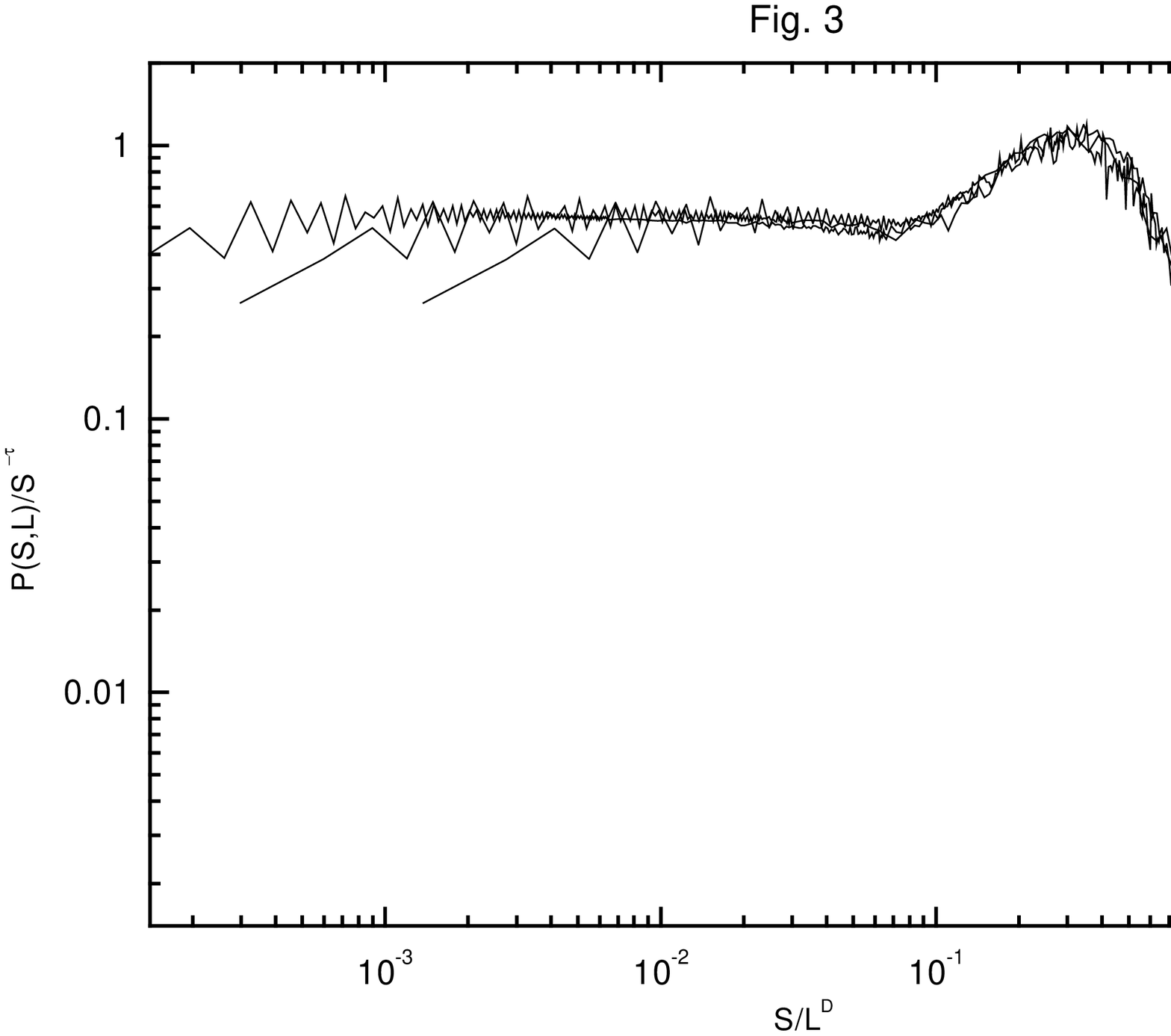,width=3in,angle=0}}
\caption{
(a) Log-Log plot of the avalanche size distributions for several values of the 
system size $L$ with $p_1=0.8$. 
(b) Data collapse of the curves displayed in (a) according to Eq. (\ref{P(s,L)}) with 
the exponents $\tau=1.53$, $D=2.2$.
}
\label{figD(s)}
\end{figure}

\begin{figure}
\centerline{\psfig{file=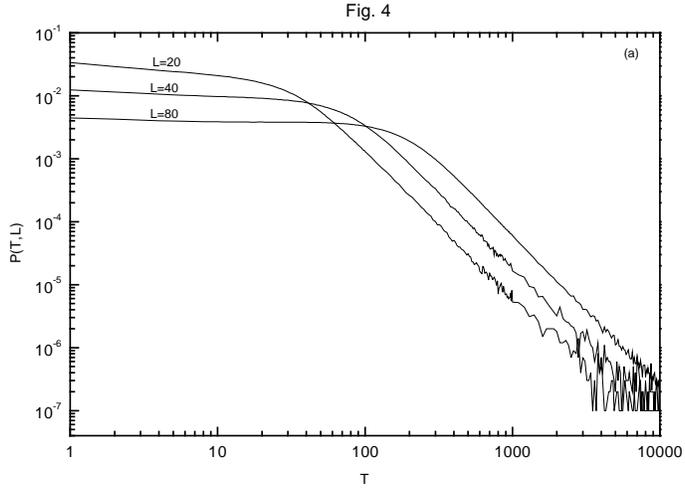,width=3in,angle=0}}
\centerline{\psfig{file=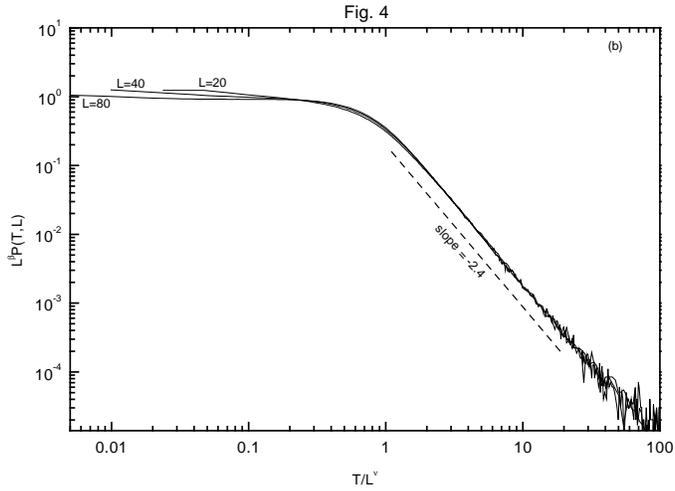,width=3in,angle=0}}
\caption{ (a) Log-Log plot of the transit-time distribution for several values 
of the system size with $p_1=0.8$.  (b) Data collapse of the curves displayed in (a) 
according to Eq. (\ref{P(T,L)}). The best fit to the numerical data gives the 
slope $\alpha=2.4$
}
\label{figP(T)}
\end{figure}

\begin{figure}
\centerline{\psfig{file=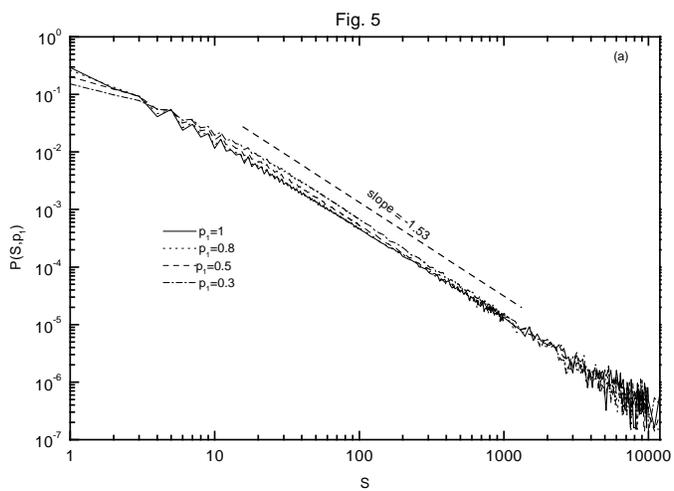,width=3in,angle=0}}
\centerline{\psfig{file=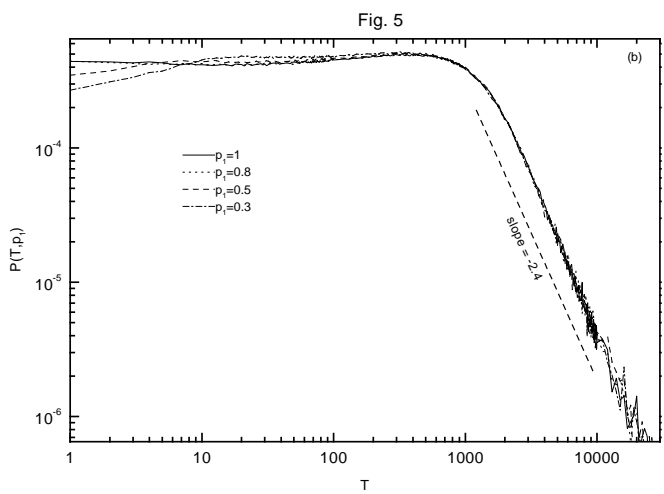,width=3in,angle=0}}
\caption{ (a) Log-Log plot of the avalanche size distribution. The best 
fit gives the slope $\tau=1.53$
(b) Log-Log plot of the transit time distribution. The best fit gives 
the slope $\alpha=2.4$
In both (a) and (b) the system size is $L=400$.  
}
\label{figDPp1}
\end{figure}

We have also studied the avalanche size and the transit time distributions 
for different values of the probability $p_{1}>p_1^c$. 
In a previous studies on the rice pile models~\cite{Amaral} it was shown 
that the avalanche size distribution is of the form:
\begin{equation}
P(S,L)=S^{-\tau}G(S/L^{D}),
\label{P(s,L)}
\end{equation}
with $\tau=1.53\pm 0.05$ and $D=2.20\pm 0.05$.
We report in Fig.~\ref{figD(s)}.a our simulation data for the distribution of 
avalanche size for different values of system size. The distribution is 
a power law with the presence of a peak close to the cutoff size 
$S_{c}\propto L^{D}$. This peak is due to the finite-size effects which leads 
the system into a supercritical state, followed by a massive avalanche. 
Since on average each site must topple exactly once during an avalanche to transport 
a grain out of the pile when the stationary state is reached, the 
average size of avalanche is thus $<S>=L$. And this leads to the scaling relation 
$D(2-\tau)=1$~\cite{Paczuski} 
The numerical results of the exponents, namely $\tau \approx 1.53$ and 
$D\approx 2.20$ are in quite good agreement with the scaling relation. 
The transit-time distribution can be  described by the scaling form (1)
 with the same exponents as the Oslo Model ($p_{1}=1$), namely, $\beta\approx 1.25$ 
and $\nu \approx 1.25$.  The power-law exponent $\alpha$ for the  large transit time is 
obtained as $\alpha\approx 2.4$. ( See Fig.~\ref{figP(T)} for an example ). 
Fig.~\ref{figDPp1}.a shows the avalanche-size distribution for a pile of size $L=400$ 
and for several 
values of jumping probability $p_{1}$. It is clear that the size exponent $\tau$
is insensitive to the value of the probability $p_{1}$ greater than a critical 
value $p_1^c$. Fig.~\ref{figDPp1}.b gives the corresponding transit time 
distribution, which is nearly a constant for small transit time, and decays as 
a power law for larger transit time. As in Fig.~\ref{figDPp1}.a, the exponents 
remain the same for several values of $p_{1}$ greater than $p_1^c$. 
Therefore, we can conclude that for $p_{1}$ greater than $p_1^c$ 
the size and transit time exponents are the same as the Oslo 
rice-pile model, thus the SOC state in our model is not affected by the 
fact that some grains topple with different jumping probability 
$p_{1}$, when $p_{1}\geq p_1^c$.   

\begin {figure}
\centerline{\psfig{file=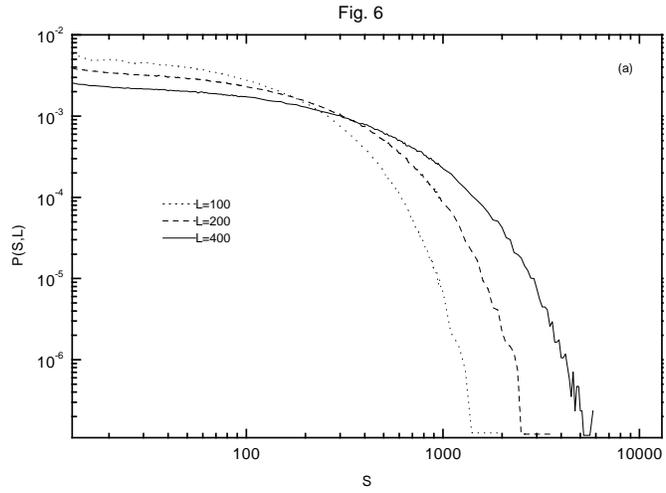,width=3in,angle=0}}
\centerline{\psfig{file=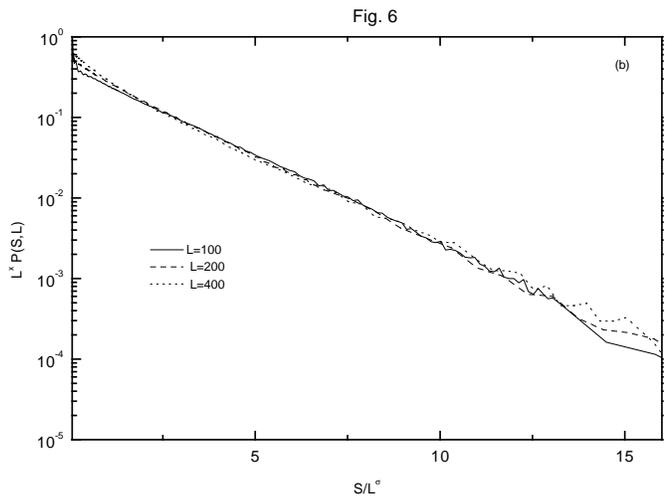,width=3in,angle=0}}
\caption{
(a) Log-Log plot of the avalanche size distribution with $p_{1}=0.01$.  
(b) Semi-log plot of the date rescaled according to Eq. (\ref{P(s,L)cross}). 
The curve is quite straight line, indicating the exponential form of the scaling 
function $G(x)$.
}
\label{figD(s)cross}
\end {figure}

\begin{figure}
\centerline{\psfig{file=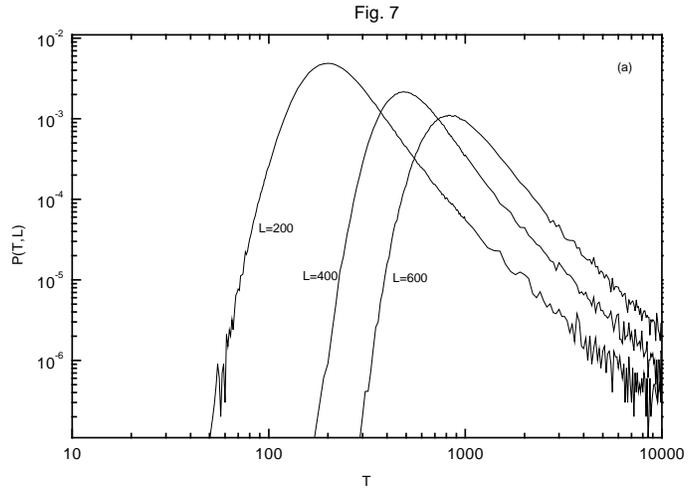,width=3in,angle=0}}
\centerline{\psfig{file=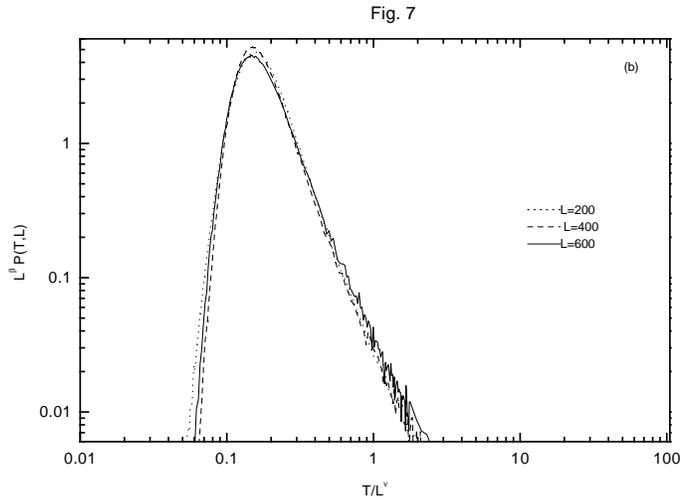,width=3in,angle=0}}
\caption{
(a) Log-Log plot of the transit-time distribution with $p_{1}=0.01$. 
(b) Log-Log plot of the rescaled data with the exponents $\beta =1.20$ and $\nu =1.20$. 
}
\label{figP(T)cross}
\end {figure}

The crossover behaviors were also investigated. We collect statistics on the 
size of the avalanche and transit time of the grains for a value of $p_{1}$ 
very close to zero ($p_{1}=10^{-2}$). As illustrated in 
Fig.~\ref{figD(s)cross}.a, 
the avalanche-size distribution does not exhibit a power-law behavior. 
The data can be described by the following form:
\begin{equation}
P(S,L)=L^{-x}G(S/L^{\sigma})
\label{P(s,L)cross}
\end{equation}  
which is confirmed by the good data collapse obtained with the exponents 
$x=0.94$ and $\sigma=0.94$ and where the scaling function $G(x)$ decays 
exponentially ( Fig.~\ref{figD(s)cross}.b ). 
The form of Eq.(\ref{P(s,L)cross}) is very different from the distribution form 
for the case $p_1=0$, which is $P(S,L)=\delta (S-L)$, where $\delta(x)$ is the Dirac 
Function. This confirms our statement that there is a sharp transition at $p_1=0$.
In Fig.~\ref{figP(T)cross}.a, 
the transit time distribution is plotted. It is visually apparent that the 
distribution peaks at the vicinity of the system size. 
A data collapse is shown in Fig.~\ref{figP(T)cross}.b, which is obtained by 
rescaling the original data according to Eq. (\ref{P(T,L)}) with $\beta =1.2$ and $\nu =1.2$. 
The exponents used here are a little smaller than that for $p_1>p_1^c$. We expect that 
when $p_1\rightarrow 0$ the exponents $\nu$ and $\beta$ shall become even smaller and finally 
approach $1$. In this case the scaling for average velocity $<v> \sim L^{-\gamma}=L^{\nu -1}$ 
will give $<v>\sim L^0=1$, which is in consistent with the results 
that $<v>=1$ for finite system at $p_1=0^+$. 


 In summary, we have investigated a one-dimensional model of a rice pile where 
the sites with higher slopes have more chance to topple (with a probability 
$p_{2}=1$) while the sites with lower slopes topple with a probability 
$p{1}\leq p_{2}$. Depending on the value of $p_{1}$, our model exhibits three 
behaviors: the trivial critical behavior for $p_{1}=0$ (1D BTW model), the crossover 
behavior due to the finite size of the rice-pile and the self-organized 
critical behaviour for $p_{1}$ greater than a certain critical value 
$p_1^c$. In fact, for a sufficiently 
large pile, $p_1^c$ goes to zero and therefore, there is a sharp 
transition between the trivial behaviour and the SOC behaviour at $p_1=0$. When the 
system exhibits the SOC behaviour, the exponents did not depend on the value 
of the probability $p_{1}$ and our model belongs to the same universality 
class as the Oslo rice-pile model. 

\centerline{\bf ACKNOWLEDGMENTS}
The authors would like to thank A. VESPIGNANI for critical readings of the 
manuscript. 
M. BENGRINE, F. MHIRECH and S. D. ZHANG would like 
to thank UNESCO, IAEA and the Abdus Salam-International Centre for Theoretical 
Physics, Trieste, Italy, for hospitality. 
S. D. ZHANG's work is supported by the National Nature Science
Foundation of China and the Educational Committee of the State
Council through the Foundation of Doctoral Training.


\end{document}